\documentclass{aastex63}

\usepackage{graphicx}
\usepackage{amsmath}
\usepackage{longtable}
\usepackage{color}
\usepackage{enumerate}

\received{XXX}
\revised{YYY}
\accepted{ZZZ}

\shorttitle{One-off and Repeating FRBs}
\shortauthors{Chen et al.}

\begin{document}

\title{One-off and Repeating Fast Radio Bursts: A Statistical Analysis}

\correspondingauthor{Wei-Min Gu; Mouyuan Sun}
\email{guwm@xmu.edu.cn; msun88@xmu.edu.cn}

\author{Hao-Yan Chen}
\affiliation{Department of Astronomy, Xiamen University, Xiamen,
Fujian 361005, P. R. China}

\author{Wei-Min Gu}
\affiliation{Department of Astronomy, Xiamen University, Xiamen,
Fujian 361005, P. R. China}

\author{Mouyuan Sun}
\affiliation{Department of Astronomy, Xiamen University, Xiamen,
Fujian 361005, P. R. China}

\author{Tuan Yi}
\affiliation{Department of Astronomy, Xiamen University, Xiamen,
Fujian 361005, P. R. China}

\begin{abstract}
According to the number of detected bursts, fast radio bursts (FRBs)
can be classified into two categories, i.e., one-off FRBs and repeating ones.
We make a statistical comparison of these two categories based on
the first FRB catalog of the Canadian Hydrogen Intensity Mapping Experiment
Fast Radio Burst Project. Using the Anderson-Darling, Kolmogrov-Smirnov, and Energy statistic tests, we find
significant statistical differences ($p$-value $<$ 0.001) of the burst
properties between the one-off FRBs and the repeating ones. 
More specifically, after controlling for distance, we find that the peak 
luminosities of one-off FRBs are, on average, higher than the repeating ones;
the pulse temporal widths of repeating FRBs are, on average, longer than the
one-off ones. The differences indicate that these two categories could have
distinct physical origins. Moreover, we discuss the sub-populations of FRBs and provide statistical evidence to support the existence of sub-populations in 
one-off FRBs and in repeating ones.

\end{abstract}


\section{Introduction} \label{sec:intro}

Fast radio bursts (hereafter FRBs) are extragalactic millisecond-duration radio
bursts and  have been detected in frequencies from 118 MHz to 8 GHz 
\citep[e.g.,][]{2019ARA&A..57..417C, Petroff et al.(2019),Pleunis et al.(2021b)}.
The host galaxies of nineteen FRBs have been identified \citep{Heintz et al.(2020)}\footnote{\url{http://frbhosts.org}}. The origins of FRBs remain 
mysteries. According to the number of detected bursts, FRBs can be classified
into two categories, i.e., repeating FRBs and observational one-off (hereafter 
one-off) ones. For repeating FRBs, models involve young rapidly rotating pulsars \citep{Lyutikov et al.(2016)}, a wandering jet in an intermediate-mass black
hole (BH) binary with chaotic accretion \citep{Katz(2017)}, a pulsar passes 
through the asteroid belt \citep{Dai et al.(2016), 2020ApJ...895L...1D},
 a neutron star (NS)-white dwarf (WD) binary with strong dipolar magnetic fields \citep{Gu et al.(2016), Gu et al.(2020)}, 
the precession of a jet in an NS/BH-WD binary with super-Eddington accretion 
rate \citep{Chen et al.(2021)}, and the intermittent collapses of the crust of a
strange star \citep{Geng et al.(2021)}.
For one-off FRBs, models often invoke the collapse of a spinning supra-massive
NS forming a BH \citep{Zhang(2014)}, and the merger of a charged Kerr-Newman BH binary \citep{Zhang(2016), Liu et al.(2016)}.
Therefore, we would expect that the repeating FRBs and one-off ones have 
different temporal and spectral properties 
\citep[e.g.,][]{CHIME/FRB Collaboration et al.(2019),Fonseca et al.(2020)}.

Recently, the Canadian Hydrogen Intensity Mapping Experiment Fast Radio Burst (hereafter CHIME/FRB) Project released the first CHIME/FRB catalog (hereafter Catalog 1); the catalog consists of FRBs that were detected between 400 and 
800 MHz and from July 25, 2018, to July 1, 2019, \citep{Amiri et al.(2021)}. 
Catalog 1 contains 492 unique sources, i.e., 474 one-off FRBs and 18 repeaters\footnote{\url{https://www.chime-frb.ca/catalog}}.
With the same search pipeline, the burst properties can be measured with uniform selection effects. Therefore, this sample is ideal for exploring the 
statistical properties of FRBs.

Some studies aim to explore the possible differences between one-off FRBs and repeating ones based on Catalog~1. 
For example, \citet{Josephy et al.(2021)} studied the Galactic latitudes of 
one-off FRBs and repeating ones and found that the distribution of FRBs is 
isotropic. Meanwhile, \citet{Amiri et al.(2021)} found that the distributions of dispersion measures (DMs) and extragalactic DMs ($\rm{DM}_E$) of one-off FRBs and the first-detected repeater events are statistically consistent with 
originating from the same underlying sample. For the distribution of fluences, \citet{Amiri et al.(2021)} showed that one-off FRBs and repeaters are 
statistically similar. On the other hand, \citet{Amiri et al.(2021)} found that
two distinct  FRB populations could be inferred from the distribution of the 
peak fluxes. Moreover, \citet{Pleunis et al.(2021a)} showed that the 
``downward-drifting'' of frequency and multiple spectral-temporal components, which are absent in one-off FRBs, seem to be common in repeaters.
Compared to the distributions of pulse temporal widths and spectral bandwidths, \citet{Pleunis et al.(2021a)} showed that the repeating FRBs (for the 
first-detected events) have larger temporal widths and narrower bandwidths than
that of one-off ones. 
\citet{Zhong et al.(2022)} showed that the distribution distinctions for one-off
FRBs and repeaters in the spectral index and the peak frequency cannot be 
explained by the selection effect due to a beamed emission.
These differences all indicate that FRBs might have different populations.
 
\citet{Amiri et al.(2021)} only compared 205 one-off FRBs and 18 the 
first-detected repeater events to explore the differences between the two types
of FRBs. The differences in the physical mechanisms of the burst features 
between repeaters and one-off FRBs could not be found only by analyzing the 
first-detected repeater events. 
Moreover, some burst parameters used in \citet{Pleunis et al.(2021a)}, e.g.,
spectral bandwidth, might be inappropriate to compare the one-off FRBs with the repeating ones. The reason is that, due to the limited instrumental bandwidth, 
the spectral bandwidth reported in Catalog 1 might be significantly biased and
is not an excellent tracer of the FRB intrinsic bandwidth. In addition, some 
previous works \citep[e.g.,][]{Cui et al.(2021),Zhong et al.(2022)} did not 
control for distance when comparing the peak luminosity and energy 
distributions of the two types of FRBs. Thus, the distinctions of comparisons
might be biased. In this work, we demonstrate that one-off FRBs and repeating
ones should be two physically distinct populations based on Catalog 1, by investigating several different burst parameters with similar distances, 
i.e., the fluxes, the pulse temporal widths, and the estimated isotropic peak
luminosity $L_{\rm p}$.

The remainder of this manuscript is organized as follows.
In Section~\ref{sec:sample}, we calculate the peak luminosities of one-off FRBs
and repeating ones, and build up the mock samples for repeaters and the control samples for one-off FRBs with matched $\rm{DM_{E}}$ to present the differences between the two types of FBRs. In Section~\ref{sec:results}, we demonstrate statistically that FRBs could be divided into two categories, 
i.e., one-off FRBs and repeaters. Then, we further discuss whether one-off FRBs 
and repeaters have sub-populations. Conclusions and discussion are presented in Section~\ref{sec:con}.

\section{The Samples} \label{sec:sample}

Our study utilizes the largest FRB sample to date (i.e., Catalog 1).
Catalog 1 consists of 474 one-off FRBs and 18 repeating FRBs (62 bursts). We 
exclude six one-off sources that have bad gains \citep{Amiri et al.(2021)}.  
Thus our sample contains 468 one-off sources plus 18 repeating sources.

\subsection{Isotropic Peak Luminosity} \label{subsec: L&E}

\begin{figure}[htbp]
\centering
\includegraphics[height=10cm,width=10cm]{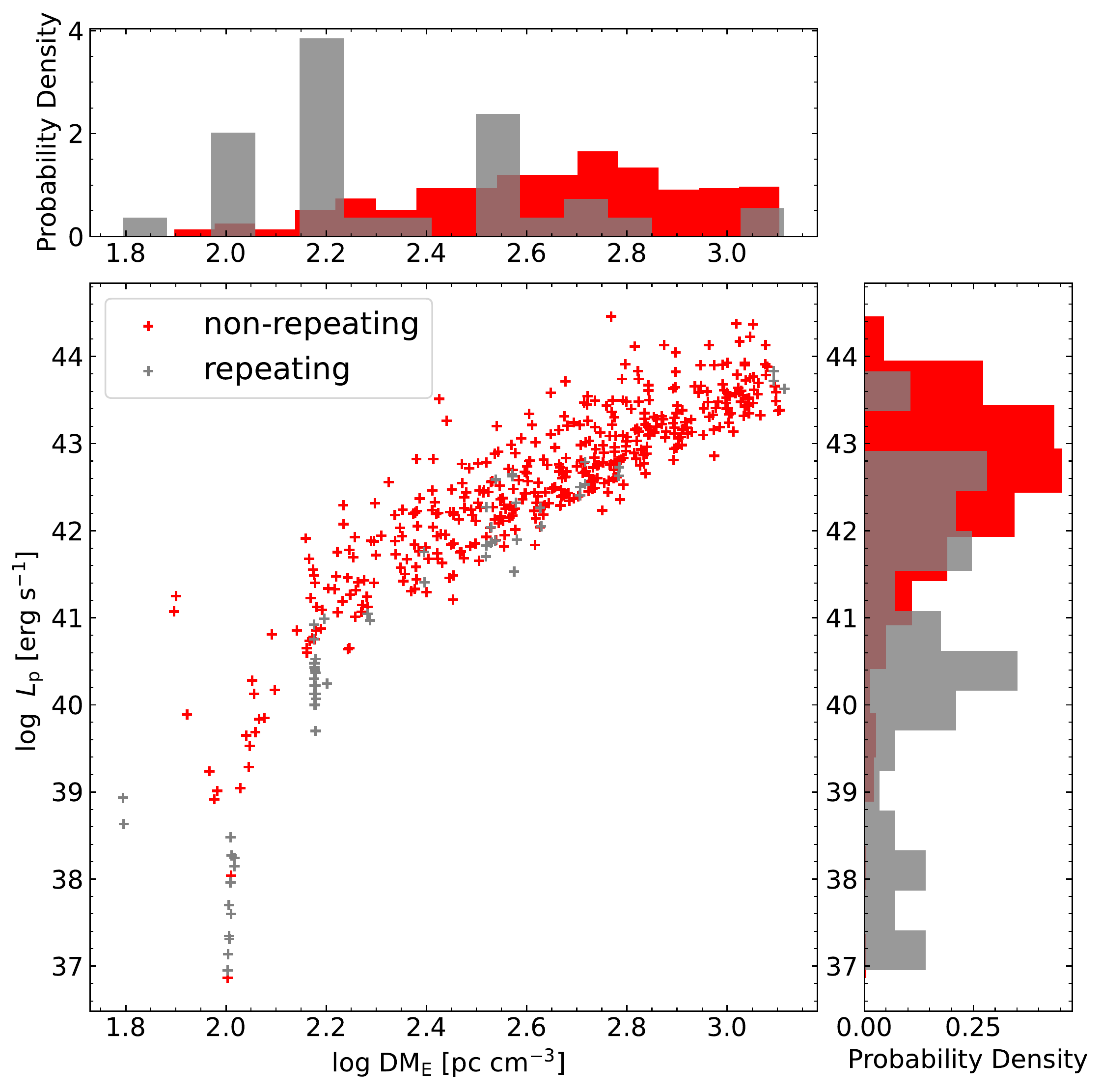}
\caption{The distributions of $\rm {DM_{\rm{E}}}$ versus $L_{\rm{p}}$ for 
one-off FRBs (red) and repeating FRBs (grey) from Catalog 1. We compare the distributions under the same upper limit of $\rm {DM_{\rm{E}}}$ 
(or redshifts $z$). The top and right distributions are marginal probability 
density estimations of the one-dimensional distribution of the corresponding parameters, i.e., $\rm{DM}_{\rm{E}}$ and $L_{\rm{p}}$. Using the 2DKS test, 
we can infer the correlation between one-off FRBs and repeaters. There could be evidence that the 2D distributions for one-off FRBs and repeating ones are inconsistent with originating from the same underlying distribution, 
i.e., $p_{\rm 2DKS} \approx 1.18 \times 10^{-10}$. For the E-statistic, the 
energy distance of $\sim 75$ ($p_{\rm{Energy}}\sim 5\times 10^{-4}$) confirms
the statistical differences between the 2D distributions.
\label{fig:1} }
\end{figure}

To estimate the FRB isotropic peak luminosity (denoted as $L_{\rm p}$), we use
FRB DMs to estimate their redshifts $z$ and luminosity distances $D_{\rm L}$
since only three host galaxies of FRBs in Catalog 1 (i.e., FRBs 20121102A, 
20180916B, and 20181030A) have been identified \citep[e.g.,][]{Zhang(2018), Macquart et al.(2020)}. The methodology is as follows. 

The extragalactic dispersion measure $\rm{DM_{E}}$ of an FRB, which takes away the contributions from our Galactic interstellar medium (ISM) ($\rm{DM}_{\rm{MW, ISM}}$) based on the Galactic electron density model NE2001 \citep{2002astro.ph..7156C} or YMW16 \citep{Yao et al.(2017)} and halo ($\rm{DM}_{\rm{MW, halo}}$), where $\rm{DM}_{\rm{MW, halo}}\approx 50\ \rm{pc}\ \rm{cm}^{-3}$ \citep{Macquart et al.(2020)}, can be written as

\begin{equation}
\rm{DM}_{\rm{E}}=\rm{DM}_{\rm{IGM}}+\rm{DM}_{\rm{host}}\ ,
\label{con:DMe}
\end{equation}
where $\rm{DM}_{\rm{IGM}}$ is the contribution from the intergalactic medium (IGM), and $\rm{DM_{host}}$ is the contribution from the host galaxy of FRB.
\citet{Amiri et al.(2021)} mainly compared the distribution of $\rm{DM}_{\rm{E}}$ based on the NE2001 model. Thus, we only consider the NE2001 model.
The $\rm{DM}_{\rm{host}}$ can be estimated as ${\rm DM_{host}}=50/(1+z)\ \rm{pc \ cm}^{-3}$ \citep{Macquart et al.(2020)}. The contribution from IGM is roughly proportional to the redshift $z$ of the source through \citep{2014ApJ...783L..35D}

\begin{equation}
{\rm DM_{IGM}}=\frac{3c H_{0} \Omega_{b} f_{\rm{IGM}}}{8 \pi G m_{p}} \int_{0}^{z}\ \frac{\chi(z) (1+z)}{\left[\Omega_{m} (1+z)^{3}+ \Omega_{\Lambda} \right]^{1/2}} \ dz \ ,
\label{con: DM_IGM}
\end{equation}
where $G$ is the gravitational constant, $f_{\rm{IGM}}$ is the fraction of
baryons in the IGM with the average value $f_{\rm{IGM}} \sim 0.83$ 
\citep{Fukugita et al.(1998)}, $m_{p}$ is the mass of proton, 
$\chi(z)=(3/4)y_{1} \chi_{e, {\rm H}}(z)+(1/8)y_{2}\chi_{e, {\rm He}}(z)$ 
expresses the free electron number per baryon in the universe, 
$y_{1} \sim y_{2} \sim 1$ are the hydrogen and helium mass fractions 
normalized to 3/4 and 1/4, respectively, and $\chi_{e, {\rm H}}$ and 
$\chi_{e, {\rm He}}$ are the ionization fractions for hydrogen and helium, respectively. For FRBs at $z<3$, both hydrogen and helium are fully ionized \citep[e.g.,][]{2016ApJ...830L..31Y}. 
Thus, $\chi_{e, {\rm H}}=\chi_{e, {\rm He}}=1$ and $\chi(z) \simeq 7/8$. 
The values of cosmological parameters are adopted from the latest Planck results \citep{Planck Collaboration et al.(2016)}, i.e., the Hubble constant 
$H_{0}\simeq 67.74\ \rm{km} \ s^{-1} \ \rm{Mpc}^{-1}$, $\Omega_{b}$  $\simeq 0.0486$, $\Omega_{m}$  $\simeq 0.3089$, and $\Omega_{\Lambda}$ $\simeq 0.6911$. Based on the result of \citet{Zhang(2018)}, the ${\rm DM_ {IGM}}$ 
can be roughly estimated when $z<3$

\begin{equation}
{\rm DM_{ IGM} }\sim z \ 855 \ \rm{pc \ cm}^{-3}\ .
\label{con:DMigm}
\end{equation}
In summary, we can use Equations (\ref{con:DMe})-(\ref{con:DMigm}) to estimate $\rm {DM_ {IGM}}$ and redshift $z$ (and the corresponding luminosity distance $D_{\rm L}$). 

The isotropic peak luminosity $L_{\rm{p}}$ of each one-off FRB or every burst of each repeating one is \citep{Zhang(2018)}

\begin{equation}
L_{\rm{p}} \backsimeq 4 \pi \ \left(\frac{D_{\rm{L}}}{\rm{10^{28}\ \rm{cm}}}\right)^{2}\ \left(\frac{S_{\nu}}{\rm{Jy}}\right) \ \left(\frac{\nu_{\rm{c}}}{\rm{GHz}}\right) \  \times 10^{42} \ \rm{erg}\ \rm{s}^{-1}\ ,
\label{con:L}
\end{equation}
where $S_{\nu}$ is the observed peak flux density of an FRB, and $\nu_{\rm{c}}$
is the central frequency of the telescope, i.e., for CHIME, $\nu_{\rm{c}}=600\ \rm{MHz}$.

We present in Figure \ref{fig:1}, the distributions of ${\rm DM_{E}}$ and 
$L_{\rm p}$ (and their joint distribution) for one-off FRBs (red) and repeating
ones (grey). We compare the distributions of ${\rm DM_{E}}$ and $L_{\rm p}$ under the same upper limit of ${\rm DM_{E}}$. The peaks of the $\rm{DM}_{\rm{E}}$ and $L_{\rm p}$ distributions of repeaters tend to be
smaller and lower than that of one-off ones, respectively.
Using the two-dimensional Kolmogrov-Smirnov (2DKS) test\footnote{\url{https://github.com/syrte/ndtest/blob/master/ndtest.py}} \citep{Peacock(1983),1987MNRAS.225..155F}, we compare the joint distribution of ${\rm DM_{E}}$ and $L_{\rm p}$  for one-off FRBs with that for repeating ones; 
we find that the differences between the two joint distributions are 
statistically significant since the corresponding $p_{\rm 2DKS}$-value is 
$\sim 1.18\times 10^{-10}$.
In addition to the 2DKS test, we use the Energy statistic
\footnote{The  E-statistic can be available for R on the Comprehensive R Archive Network (CRAN) under general public license, i.e., $energy$ package (\url{https://cran.r-project.org/web/packages/energy/index.html}).}
\citep[E-statistic; e.g.,][]{Gabor et al.(2013)} for two-dimensional data to 
verify the differences in the two joint distributions of one-off FRBs and 
repeating ones. The E-statistic test compares the energy distance between the distributions of the different samples. The energy distance is zero if and only 
if the distributions are identical. The energy distance of two joint 
distributions for one-off FRBs and repeating ones is $\sim 75$ 
($p_{\rm{Energy}}\sim 5\times 10^{-4}$), implying that the two joint 
distributions of ${\rm DM_{E}}$ and $L_{\rm p}$ are significantly different. 
The observed differences suffer from strong statistical biases, since the two
types of FRBs have different $\rm{DM_{E}}$ distributions. Below, we explore
this point in detail.

\subsection{The  Mock Samples} \label{subsec: new}

As shown in Figure~\ref{fig:1}, one-off FRBs and repeating ones have different distributions of $L_{\rm p}$.
The differences might be intrinsic, i.e., one-off FRBs and repeaters have 
distinct bursting energy generation mechanisms, or caused by the differences in redshift $z$ (or $D_{\rm L}$). Hence, we perform the following Monte Carlo experiment to build the mock samples of repeaters and one-off FRBs with the matched ${\rm DM_{E}}$ to verify whether the differences are intrinsic. Unlike previous works, which only chose the physical 
properties of the first-detected burst for each repeater to explore the 
differences between the two types of FRBs 
\citep[e.g.,][]{Amiri et al.(2021),Pleunis et al.(2021a)}, we consider each 
repeating burst as an independent mock FRB source. The reason is that, as shown
in Figure~\ref{fig:1}, the peak luminosities of different bursts from the same 
repeater vary greatly (e.g., $\sim 1.5$ magnitudes for FRB~20180916B). The burst features of the repeaters could not be revealed only by analyzing the 
first-detected bursts. We expect that, if the physical mechanisms for one-off 
FRBs and repeating ones are the same, the peak luminosity distribution of all
bursts in repeating FRBs is statistically similar to bursts in one-off ones
after controlling $\rm{DM_{E}}$. We conduct Monte Carlo experiments to test
 the expectation. In the Monte Carlo experiments, we need to ensure that each 
burst of repeaters has a corresponding one-off FRB with a similar 
$\rm{DM_{E}}$, i.e., $\vert \rm{DM_{repeating}}-\rm{DM_{one-off}}\vert/\rm{DM_{repeating}} \la 0.05$. For the repeaters, the minimum
value of the $\rm{DM_{E}}$ is $62.5\ \rm{pc\ cm^{-3}}$ of FRB~20181030A. None of the one-off FRBs has the similar $\rm{DM_{E}}$ as FRB~20181030A since the minimum value of the one-off FRB $\rm{DM_{E}}$ is $78.8\ \rm{pc\ cm^{-3}}$. Thus, for the Monte Carlo experiments, we reject the repeater FRB 20181030A and consider the remaining 17 repeaters which have a total number of 60 bursts. 

\begin{figure*}[htbp]
\centering
\includegraphics[height=8cm,width=16cm]{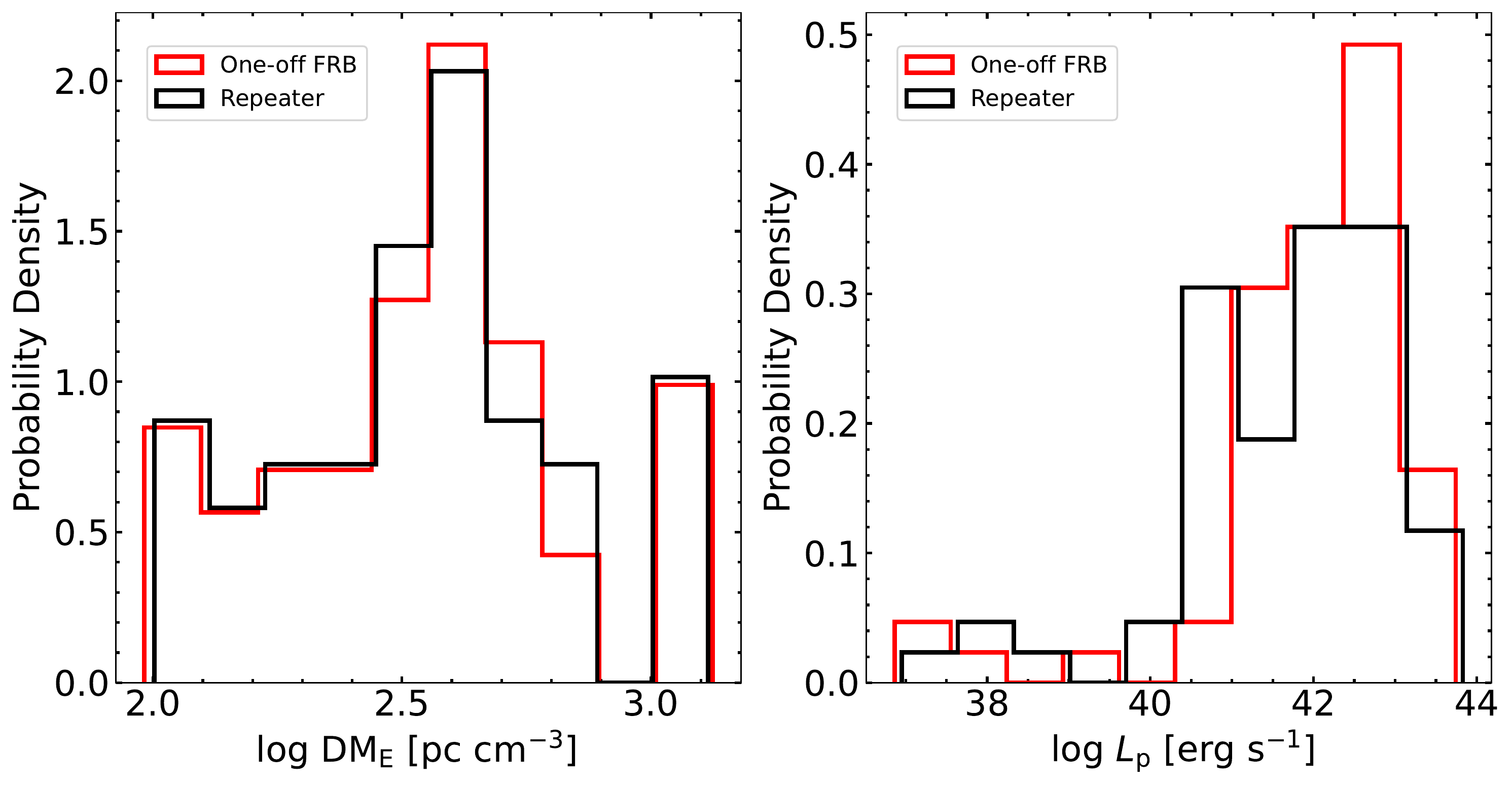}
\caption{The probability density distributions of $\rm{DM_{E}}$ 
(the left panel) and $L_{\rm{p}}$ (the right panel) for a random realization of 
the repeating mock and the one-off control samples. For the distribution of $\rm{DM_{E}}$, two samples should arise from the same group, i.e., $p_{\rm{KS}} \approx 0.95$. For the distribution of $L_{\rm{p}}$, two mock samples could originate from different underlying distributions, since $p_{\rm{KS}} \approx 
4.16 \times 10^{-6}$.
\label{fig2}}
\end{figure*}

The steps to conduct the Monte Carlo experiments are as follows.
First, we randomly select (with replacement) 60 mock repeating FRBs from the 17 repeaters. Hence, each repeater can be selected multiple times 
($n_\mathrm{r}$), and the statistical expectation of $n_\mathrm{r}$ is the same
for all repeaters. Second, for each mock repeating FRB, we randomly select (with replacement) one burst, whose DM is the same as the mock repeating FRB. That
is, we eventually obtain 60 bursts (hereafter repeating mock sample) from the repeating FRBs, and the contribution of each repeater to the mock sample is statistically the same. 
The DM distribution of the repeating mock sample is different from that of the original 60 bursts mentioned in Section~\ref{subsec: L&E} and 
Figure~\ref{fig:1}, but resembles the DM distribution of the 17 repeaters. 
Third, for each repeating mock burst, we randomly select a one-off FRB with the similar ${\rm DM_{E}}$ (hereafter the one-off control sample).
Hence, the repeating mock sample and the one-off control sample share identical $\rm{DM_E}$ distributions (see the left panel of Figure~\ref{fig2}). 
Then, we compare the peak luminosity distributions of the repeating mock sample and the one-off control sample (see the right panel of Figure~\ref{fig2}). We 
repeat this experiment ten thousand times. The mean distributions (and their $2\sigma$ uncertainties) of the peak luminosities $L_{\rm{p}}$ of bursts
from the repeating mock samples and one-off control samples are calculated.

\begin{figure*}[htbp]
\centering
\includegraphics[height=8cm,width=16cm]{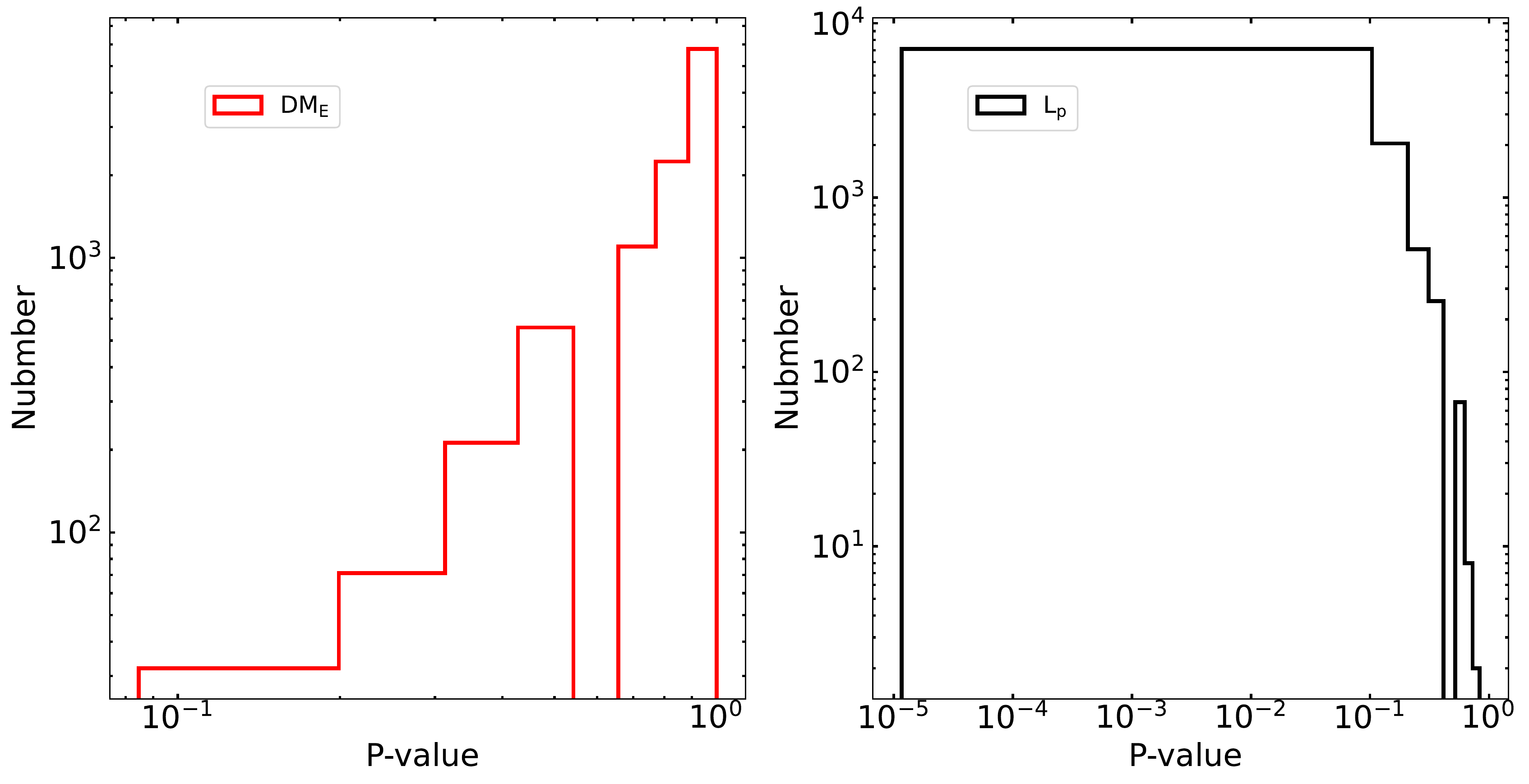}
\caption{Left: the $p$-value of the K-S test for the $\rm{DM_{E}}$ distributions of the two mock samples in the ten thousand simulations. 
Right: the same as the left panel, but for $L_{\rm{p}}$. For all simulations, 
the $p$-value for $\rm{DM_{E}}$ is always larger than 0.05. For $90\%$ of simulations, the $p$-value for $L_{\rm{p}}$ is smaller than 0.05. 
\label{fig3}}
\end{figure*}

We use the Kolmogrov-Smirnov (KS) test to assess whether there are statistical differences between the repeating mock samples and the one-off 
control samples. The two samples are statistically different if the $p$-values 
of the KS tests are smaller than $5\%$. The probability density distributions 
of the $\rm{DM_{E}}$ and $L_{\rm{p}}$ for a realization of the repeating mock
and the one-off control samples are shown in Figure~\ref{fig2}. For the 
distributions of $\rm{DM_{E}}$ (the left panel of Figure~\ref{fig2}), the 
$p$-value of the KS test is $\sim 0.95$. For the distributions of $L_{\rm{p}}$
(the right panel of Figure~\ref{fig2}), the $p$-value of the K-S test is 
$\sim 4.16 \times 10^{-6}$, indicating that two mock samples could originate
from different underlying peak-luminosity distributions after controlling for $\rm{DM_{E}}$. For the ten thousand simulations, while the $p$-value of the
K-S test for the $\rm{DM_{E}}$ distributions (the left panel of 
Figure~\ref{fig3}) is always larger than 0.05, the $p$-value for the 
$L_{\rm{p}}$ distributions (the right panel of Figure~\ref{fig3}) is smaller
than 0.05 in $90\%$ of  simulations. 
We also use the E-statistic test to check the statistical differences in the 
distributions of $\rm{DM_{E}}$ and $L_{\rm{p}}$ and obtain the same conclusions. 

Some studies have found the differences in the temporal widths between repeating FRBs and one-off ones \citep[e.g.,][]{Amiri et al.(2021),Pleunis et al.(2021a)}.
Therefore, in addition to the comparison of the luminosity distributions,
we follow the above procedures to compare the distributions of the pulse 
temporal widths of repeating FRBs and one-off ones. The mean distributions 
and the corresponding $2\sigma$ statistical uncertainties are evaluated.
In our comparisons, after controlling for ${\rm DM_{E}}$, the pulse temporal 
widths of two types of FRBs would have similar dispersive delays 
\citep[e.g., Equation (4) in][]{Petroff et al.(2019)}, which could help to 
compare the distribution distinctions for the intrinsic pulse temporal widths 
between one-off FRBs and repeaters.
For the peak flux density $S_{\nu}$, there is a correlation between $L_{\rm{p}}$ and $S_{\nu}$ (see Equation (\ref{con:L})).
Hence, we mainly compare the differences in $L_{\rm{p}}$ and the burst temporal widths between repeating FRBs and one-off ones. For the distributions 
of the temporal widths, there are statistical differences between the repeating
mock samples and the one-off control samples since the $p$-values of the KS
and E-statistic tests are always less than 0.001 in all simulations.

\section{The populations of FRBs} \label{sec:results}

\subsection{Repeating and one-off FRBs} \label{subsec: r and non}

\begin{figure*}
\centering
\includegraphics[height=8cm,width=16cm]{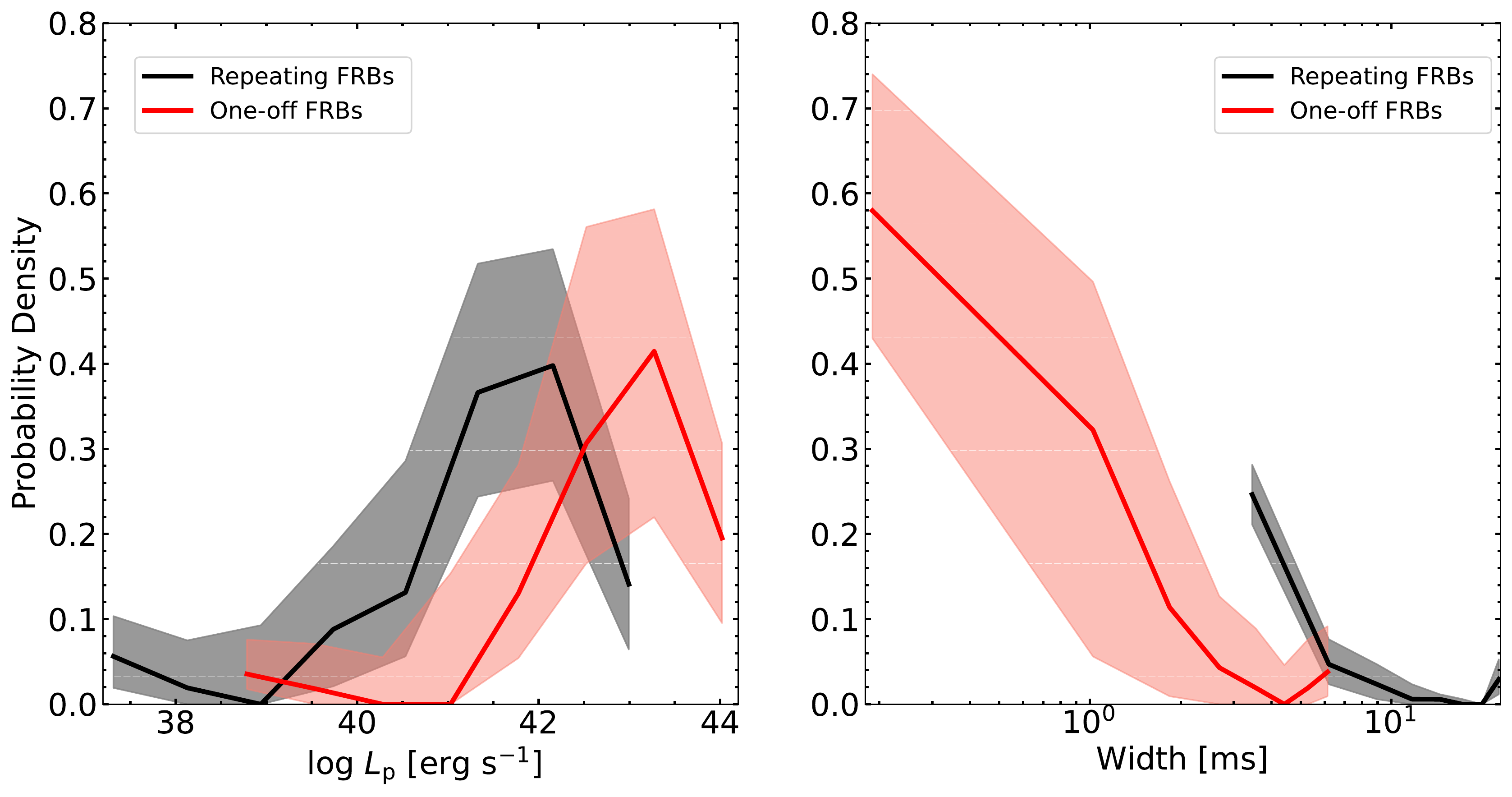}
\caption{The probability density distributions of $L_{\rm{p}}$ (left) and 
intrinsic temporal widths (right) for one-off FRBs and repeating ones. In both
panels, the red lines show the mean distributions of one-off control samples and
the black lines represent the mean distributions of repeating mock samples. All shadow areas indicate the two-sigma ($95\%$) confidence intervals. 
The distributions of $L_{\rm{p}}$ show that the peak luminosities of one-off 
sources are, on average, higher than the repeating ones. The distributions are inconsistent with arising from the same FRB group (i.e., $p_{\rm AD}<0.001$, $p_{\rm KS}\approx 2\times 10^{-6}$). Furthermore, as shown by the 
distributions of temporal widths, the pulse temporal widths of repeating FRBs 
are, on average, longer than the one-off ones. There is evidence that two FRB populations might originate from different underlying distributions, 
i.e., $p_{\rm AD}<0.001$ and $p_{\rm KS}\approx 2.5\times 10^{-5}$.
\label{fig:4}}
\end{figure*}

Figure \ref{fig:4} shows the mean distributions of $L_{\rm{p}}$ 
(the left panel) and the pulse temporal widths (the right panel) for the bursts
of repeating FRBs and one-off ones with similar $\rm{DM_{E}}$. Their corresponding $2\sigma$ statistical uncertainties are shown as the shaded 
regions. We use the Anderson-Darling (AD) and KS tests to assess the
statistical differences in two FRB samples in terms of $L_{\rm{p}}$ and the
temporal width. The null hypotheses of the AD and KS tests are that the two 
samples are drawn from the same population. If the $p$-values of the AD and 
KS tests are smaller than $5\%$, we reject the null hypothesis and conclude that
the two samples are statistically different. For the distributions of $L_{\rm{p}}$ (the left panel of Figure~\ref{fig:4}), we find that the 
$p$-values of KS and AD tests are less than 0.001. That is, the bursts of the
two FRB samples have intrinsically different luminosity distributions (the mean $L_{\rm{p}}$ for repeaters is log$\left(L_{\rm{re}}/\rm{erg \ s^{-1}}\right) 
\simeq 40.14 \pm 0.71$; for one-off FRBs is log$\left(L_{\rm{one}}/\rm{erg \ s^{-1}}\right) \simeq 41.41 \pm 0.65$) and the difference is more evident at 
the low luminosity end (log$\left(L_{\rm{re}}/\rm{erg \ s^{-1}}\right) \la 39$). 
Moreover, for the distributions of temporal widths (the right panel of Figure~\ref{fig:4}), we find that the $p$-values of KS and AD tests are 
smaller than 0.001. That is, the bursts of the two FRB samples have distinct distributions of the temporal width (the mean temporal width for repeaters 
is $13.08 \pm 2.38 \ \rm{ms}$; for one-off ones is $3.14 \pm 0.74 \ \rm{ms}$) 
and the difference is more evident at the long temporal width end ($\gtrsim 7\ \rm{ms}$). These differences support the speculation that the emission 
mechanisms of the bursts in the two types of FRBs should be distinct.

\subsection{Sub-populations} \label{subsec: sub}

\begin{figure}
\centering
\includegraphics[height=10cm,width=15cm]{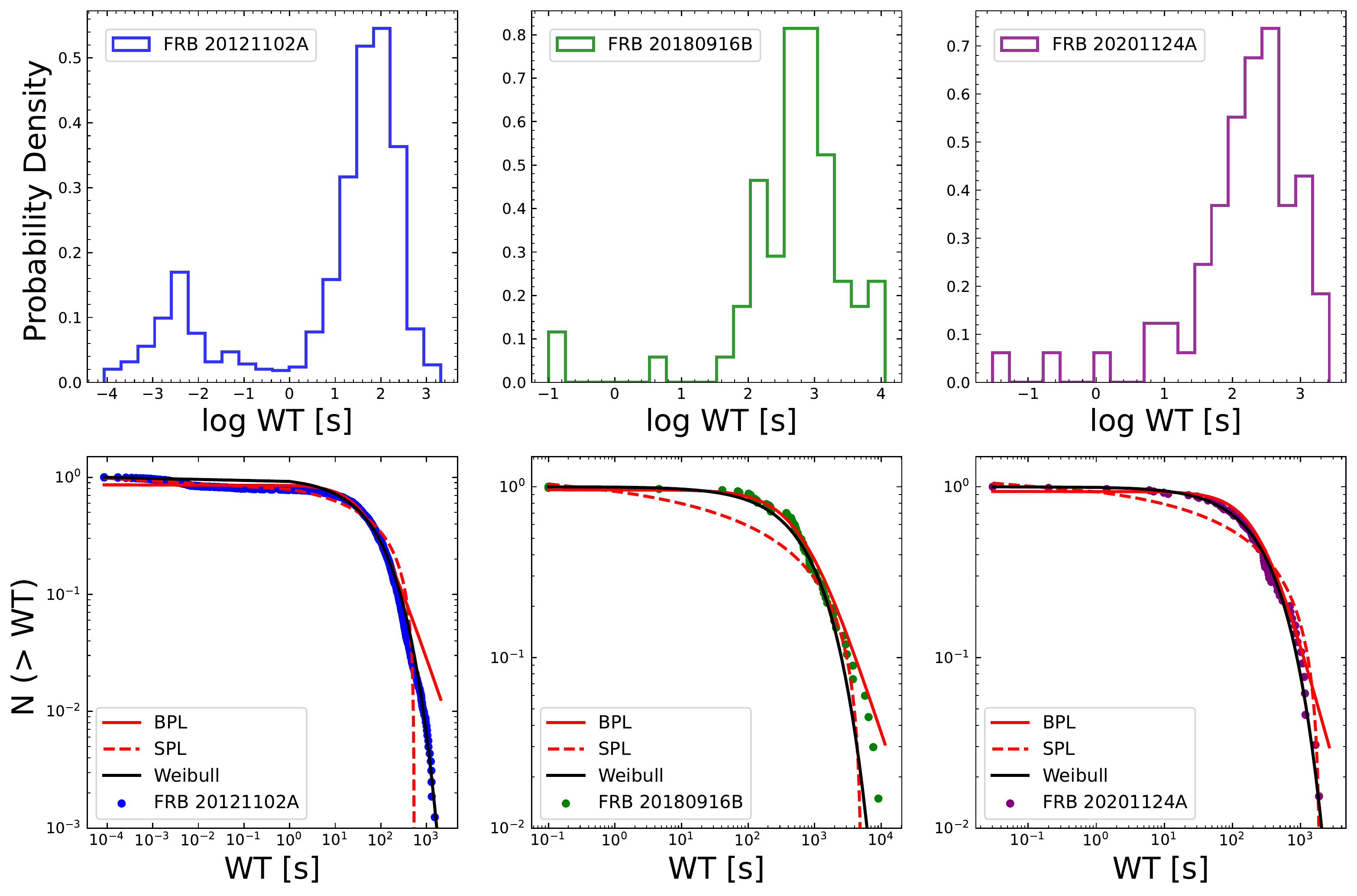}
\caption{Top: the distributions of waiting times (WTs) of three active repeating
FRBs. The WT distributions of FRB 20121102A show a bimodal structure, whereas other FRBs, i.e., FRBs 20180916B and 20201124A, have unimodal structures.
Bottom: the cumulative distribution functions (CDFs) of WTs of three FRBs. Three distribution models, i.e., the bent power law (BPL; red line), the simple power
law (SPL; dashed red line), and Weibull model (black line), are considered to
fit the CDFs of WT. The data for FRB 20121102A is taken from 
\citet{Li et al.(2021)}; the data for FRB 20180916B is taken from 
\citet{Chawla et al.(2020)}, \citet{CHIME/FRB Collaboration et al.(2020)}, \citet{Marthi et al.(2020)}, \citet{Pastor-Marazuela et al.(2021)}, and 
\citet{Pleunis et al.(2021b)};
the data for FRB 20201124A is taken from \citet{Hilmarsson et al.(2021)}, and \citet{Marthi et al.(2022)}. 
\label{fig5} }
\end{figure}

As mentioned in Section \ref{subsec: r and non}, repeating FRBs and one-off ones should have distinct emission mechanisms.  
Do one-off FRBs and repeating FRBs have sub-populations? 

For one-off FRBs, the single-pulse bursts are commonly detected \citep[e.g.,][]{CHIME/FRB Collaboration et al.(2019)}, whereas
\citet{CHIME/FRB Collaboration et al.(2021)} detected the multi-component pulse profiles of  three one-off FRBs, i.e., FRBs 20191221A, 20210206A, and 20210213A, 
and identified the periodic separations of these components. The complex and periodic profiles indicate that there could exist a sub-group of one-off FRBs 
which have similar physical processes or host environments to some repeaters. 

For periodicity, two repeating sources have been found periodic activities,
i.e., FRBs 20180916B \citep[16.35-day activity period; see][]{CHIME/FRB Collaboration et al.(2020)} and 20121102A 
\citep[$\sim$157-day activity period; see][]{Cruces et al.(2021)}. 
Some repeaters with more than 20 bursts recorded by CHIME were not found to
have any periodic activity, e.g., FRBs 20180814A (22 bursts), 20190303A
(27 bursts), and 20201124A (34 bursts). The differences in periodicity of 
repeating FRBs might indicate that the repeaters could have different groups, 
i.e., the periodic repeaters and the aperiodic repeaters 
\citep[e.g.,][]{Katz(2017),Lin et al.(2022)}.  

For active repeaters, we can study the waiting time (WTs) distributions to 
understand the physical mechanism of repeating FRBs \citep[e.g.,][]{2017JCAP...03..023W,Zhang et al.(2021)}. The WTs are the time intervals
between two adjacent (detected) bursts measured during periods of continuous observation. Up till now, the most active repeaters are FRBs 20121102A
\citep[e.g.,][]{Li et al.(2021)}, 20180916B 
\citep[e.g.,][]{CHIME/FRB Collaboration et al.(2020)} and 20201124A 
\citep[e.g.,][]{Xu et al.(2021)}.
As shown in the top three panels of Figure \ref{fig5}, the distributions of WTs
for three active repeating FRBs have different features. i.e., a bimodal
structure for FRB 20121102A, and unimodal structures for FRBs 20180916B and 20201124A. We can infer some progenitor models for repeaters from the
cumulative distributions functions (CDFs) of WTs. Three models are proposed 
to fit the CDFs of WTs, i.e., simple power law (SPL) \citep[e.g., the soft gamma repeaters (SGRs)-like model for FRB 121102;][]{2017JCAP...03..023W}, bent power law (BPL) \citep[e.g., SGR J1550-5418;][]{Chang et al.(2017)}, and Weibull
function \citep[e.g.,][]{Oppermann et al.(2018)}. The SPL model is expressed as \citep{2021MNRAS.tmp.3280S}

\begin{equation}
N(> \delta t)=A\left(\delta t^{-\alpha}-t_{\rm c}^{-\alpha}\right)\ ,
\label{con:SPL}
\end{equation}
where $\delta t$ is the waiting time and $t_{\rm c}$ is the cut-off value with $N(>t_{\rm c})=0$.
The BPL model is taken the form \citep{2021MNRAS.tmp.3280S}

\begin{equation}
N(> \delta t)=B\left[1+\left(\frac{\delta t}{t_{\rm m}}\right)^{\beta}\right]^{-1}\ ,
\label{con:BPL}
\end{equation}
where $t_{\rm m}$ is the median value, i.e., the number of bursts with 
$\delta t > t_{\rm m}$ is equal to the number of bursts with 
$\delta t < t_{\rm m}$. The Weibull function is described as \citep[e.g.,][]{Oppermann et al.(2018)}

\begin{equation}
N(> \delta t)= e^{-(\delta{t} {r} \Gamma(1+1/k))^{k}}\ ,
\label{con:Weibull}
\end{equation}
where $k$ is the shape parameter, $r$ is the event rate, and $\Gamma$ is the
gamma function. The Weibull distribution is equivalent to a
Poisson distribution when $k$ = 1, while $k\sim1$ implies that the bursts
are clustered, with more clustering implied for lower $k$.
The fitting results are shown in the bottom three panels of Figure \ref{fig5}. 
The best-fitting parameters and their $1\sigma$ uncertainties for three
fitting models are shown in Table~\ref{table:1}. The $\chi^{2}$, defined as $\chi^{2}=\sum [(N_{i}-N_{i,\rm{model}})^{2}/N_{i}]$, where $N_{i}$ and
$N_{i,\rm{model}}$ are the data of the WT distributions and the fitting models, respectively, can be used to evaluate the goodness of fit of the fitting models 
for the CDFs of WTs.
The minimum value of $\chi^{2}$ corresponds to the best-fitting models.
We find that the CDFs of WTs for FRBs 20121102A and 20201124A can be well
fitted by the Weibull function; the WT distribution of FRB~20180916B is well 
fitted by the BPL model.
That is, the central engine of these active repeating FRBs could be magnetars, 
while the physical mechanisms should be distinct.
Using the KS test (and the AD test), we compare the correlations of the 
distributions of WTs for three active repeating FRBs.
There is evidence that three active repeaters should come from different groups, 
i.e., for FRBs 20121102A and 20180916B, $p_{\rm KS}=0$; for FRBs 20121102A and 20201124A, $p_{\rm KS}\approx 1.67\times 10^{-10}$; for FRBs 20180916B and 20201124A, $p_{\rm KS}\approx 7.77\times 10^{-16}$.

\begin{table*}[!htpb]
   \caption{The best-fitting parameters for three CDF models}
     \label{table:1}
   \begin{center}
   \begin{tabular}{r|ccc|ccc|ccc}\hline \hline

FRB  & ~& SPL &~& ~ & BPL & ~ & ~ & Weibull  &~   \\ \hline\hline
~ &  $\alpha$  & $t_{\rm{c}}$ (s) & $\chi^{2}$ & $\beta$ &   $t_{\rm{m}}$ (s) & $\chi^{2}$ & $r$ $ (\rm{day^{-1}})$ &  $k$ & $\chi^{2}$     \\ \hline\hline
FRB 20121102A &  $-0.25^{+0.05}_{-0.05} $ & $534.91^{+200.67}_{-122.19}$ & 416.85 &  $1.14^{+0.29}_{-0.24}$ & $51.54^{+11.40}_{-9.84}$ & 8.99 &
$864.25^{+59.23}_{-43.13}$ & $0.59^{+0.12}_{-0.11}$ & 7.94 \\ \hline\hline
FRB 20180916B & $-0.16^{+0.05}_{-0.09}$ & $5060.89^{+1832.79}_{-1936.96}$ & 4.86 & $1.21^{+0.53}_{-0.58}$ & $684.99^{+214.72}_{-252.55}$ & 0.34 & $86.43^{+25.56}_{-21.84}$ & $0.78^{+0.18}_{-0.15}$ & 0.39
\\ \hline\hline
FRB 20201124A & $-0.22^{+0.06}_{-0.10}$ & $1926.37^{+716.62}_{626.42}$ & 1.28
& $1.48^{+0.34}_{-0.33}$ & $264.05^{+142.98}_{-105.77}$ & 0.52 & $241.92^{+83.68}_{-75.04}$ & $0.83^{+0.11}_{-0.10}$ & 0.14
\\ \hline
\hline\noalign{\smallskip}
  \end{tabular}
  \end{center}
\end{table*}

\begin{figure}
\centering
\includegraphics[height=6cm,width=18cm]{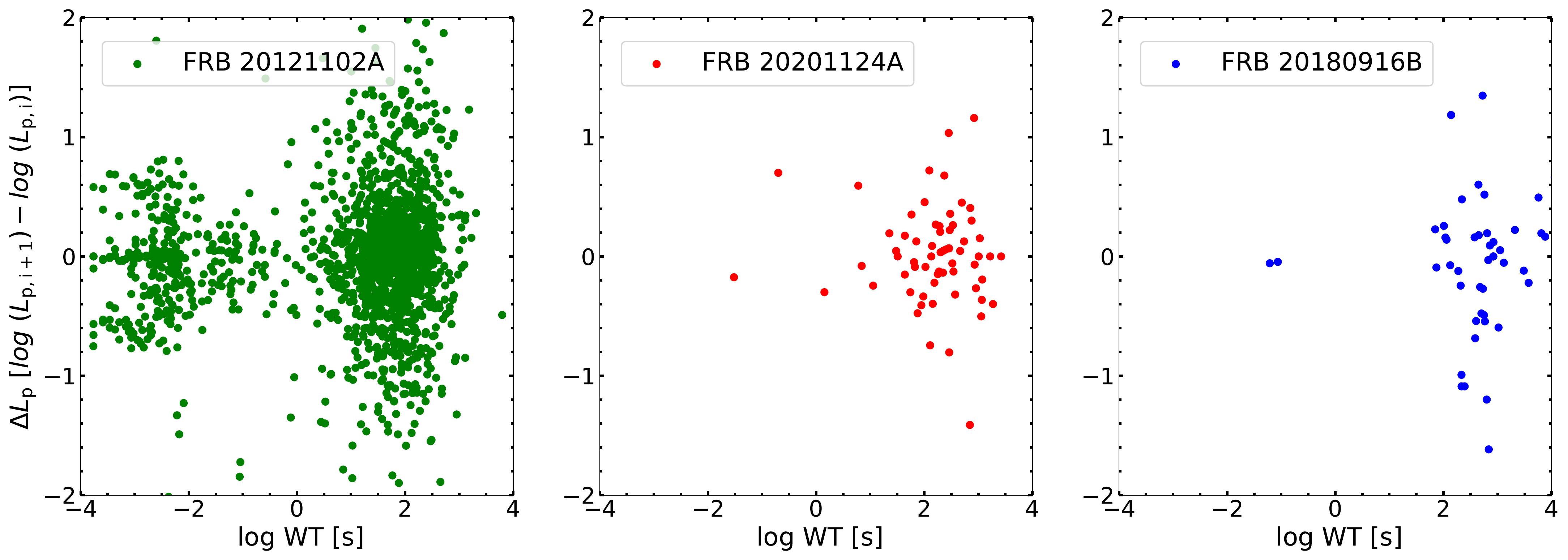}
\caption{The joint distributions of the WTs and the luminosity 
differences $\Delta L_{\rm{p}}= {\rm{log}}(L_{\rm{p, i+1}})-{\rm{log }}
(L_{\rm{p, i}})$. Three active repeaters should originate from distinct groups,
i.e., for FRBs 20121102A and 20180916B, the energy distance of $\sim 26128$ ($p_{\rm{Energy}}\sim 1\times 10^{-4}$); for FRBs 20121102A and 20201124A, 
the energy distance of $\sim 8562$ ($p_{\rm{Energy}}\sim 7\times 10^{-4}$); for FRBs 20180916B and 20201124A, the energy distance of $\sim 5719$ ($p_{\rm{Energy}}\sim 5\times 10^{-4}$).
\label{fig6} }
\end{figure}

In addition to comparing differences in the CDFs of WTs, we would like
to compare the differences between the active repeating FRBs in the joint distributions of $L_{\rm{p}}$ and WTs. As shown in Figure~\ref{fig6}, 
$\Delta L_{\rm{p}}$ represents the luminosity differences between two adjacent (detected) bursts, i.e., $\Delta L_{\rm{p}}= {\rm{log}}
(L_{\rm{p, i+1}})-{\rm{log }}(L_{\rm{p, i}})$, where $L_{\rm{p, i}}$ represents
the peak luminosity of the $i$-th burst.
We use the E-statistic test to verify the differences in the joint distributions
of three active repeaters in terms of WTs and $\Delta L_{\rm{p}}$. There is 
significant evidence that three active repeaters should
originate from distinct populations, i.e., for FRBs 20121102A and 20180916B, the energy distance of $\sim 26128$ ($p_{\rm{Energy}}\sim 1\times 10^{-4}$); 
for FRBs 20121102A and 20201124A, the energy distance of $\sim 8562$ ($p_{\rm{Energy}}\sim 7\times 10^{-4}$); for FRBs 20180916B and 20201124A, 
the energy distance of $\sim 5719$ ($p_{\rm{Energy}}\sim 5\times 10^{-4}$).

Moreover, the polarization of repeating FRBs is also a significant feature that
suggests the diversity of repeaters. FRB 20121102A was found to be $\sim 100 \%$ linearly polarized and had the rotation measure (RM) about $\sim 10^{5}$ rad $\rm{m}^{-2}$, which implies an extreme magneto-ionic environment 
\citep[e.g.,][]{Michilli et al.(2018)}. FRB 20180916B was also found to be $\sim 100\%$ linearly polarized yet is not surrounded by a dense supernova remnant
since $\left| \rm{RM} \right|\sim 10^{2}$ rad $\rm{m}^{-2}$ 
\citep{Pastor-Marazuela et al.(2021)}. 
In addition, some bursts of FRB 20201124A showed the signs of circular 
polarization, which were first found in repeating FRBs 
\citep[e.g.,][]{Hilmarsson et al.(2021)}.
The discrepancies of the distributions of WTs and the polarization
characteristics suggest that the central engine of repeating FRBs could be 
magnetars, while the emission mechanisms \citep[e.g., pulsar-like or GRB-like mechanisms; see the review of][]{Zhang(2020)} and host environments \citep[e.g., the magnetic field strengths;][]{Zhong et al.(2022)} should be distinct.

\section{Conclusions and discussion} \label{sec:con}

In this paper, we have statistically studied the differences in the burst 
properties between one-off FRBs and repeaters in Catalog 1. We found that 
one-off FRBs and repeating ones have different burst $L_{\rm p}$ and temporal
width distributions, indicating that the two samples have distinct physical 
origins (see Figure \ref{fig:4}; Section \ref{subsec: r and non}). Moreover, we 
discuss the sub-populations of one-off FRBs and repeating ones and provide a 
piece of evidence to support the existence of sub-populations in both types of 
FRBs (see Figures~\ref{fig5} and \ref{fig6}; Section \ref{subsec: sub}).

Our work can be advanced in terms of the sample size.
\citet{Chen et al.(2022)} used unsupervised machine learning to classify
repeaters and one-off FRBs from Calatog 1 and identified 188 repeating FRB candidates from 474 one-off sources, suggesting a large fraction of repeating 
FRBs could be missed due to the lack of monitoring observations. 
Recently, \citet{Li et al.(2021)} detected 1652 independent burst events from 
FRB~20121102A by FAST. The flux limit of this sample is at least three times
lower than those of previously observed samples for FRB 20121102A. Therefore,
the sensitive radio telescopes, e.g., FAST, can detect many faint bursts to 
help to reveal the physical nature of FRBs and to determine whether the distinct populations of FRBs are present.

\acknowledgments

We thank the scientific and statistic referees for helpful comments
that significantly improved the paper. This work was supported by
the National Key R\&D Program of China under grant 2021YFA1600401,
and the National Natural Science Foundation of China under grants 11925301,
11973002, 12033006, and 12221003.

\end{document}